11

**Consciousness beyond neural fields: expanding the possibilities of what has not yet happened**

Birgitta Dresp-Langley

*Centre National de la Recherche Scientifique UMR 7357 CNRS, Strasbourg University, Strasbourg Cedex, France*<u>**Cite as**</u> **:** Dresp-Langley B (2022). Consciousness Beyond Neural Fields: Expanding the Possibilities of What Has Not Yet Happened. *Frontiers in Psychology*. **12**:762349.

**Abstract**

In the field theories in physics, any particular region of the presumed space-time continuum and all interactions between elementary objects therein can be objectively measured and/or accounted for mathematically. Since this does not apply to any of the field theories, or any other neural theory, of consciousness, their explanatory power is limited. As discussed in detail herein, the matter is complicated further by the facts than any scientifically operational definition of consciousness is inevitably partial, and that the phenomenon has no spatial dimensionality. Under the light of insights from research on meditation and expanded consciousness, chronic pain syndrome, healthy ageing, and eudaimonic well-being, we may conceive consciousness as a source of potential energy that has no clearly defined spatial dimensionality, but can produce significant changes in others and in the world, observable in terms of changes in time. It is argued that consciousness may have evolved to enable the human species to generate such changes in order to cope with unprecedented and/or unpredictable adversity. Such coping could, ultimately, include the conscious planning of our own extinction when survival on the planet is no longer an acceptable option.

**Introduction**

In field theories of consciousness (e.g. Köhler, 1940; Cacha and Poznanski, 2014 among others), the latter is conceived in terms of a field in the sense in which it is used in quantum or particle physics, where the notion of 'field' applies to all fundamental forces and relationships between elementary particles within a unifying theoretical framework where the forces lead to energy fields that occupy space-time and mediate interactions between elementary particles. In field theories of consciousness the latter is, similarly, seen as having duration and extension in space. In field theories in physics, however, each point of a particular region of the presumed space-time continuum, as well as all interactions between elementary objects, are objectively measurable and accounted for mathematically. This cannot be claimed by any current theory of consciousness, including the field theories (e.g. Köhler, 1940; Lashley, Chow, and Semmes, 1951; John, 2002; McFadden, 2002; Fingelkurts and Fingelkurts, 2002; Cacha and Poznanski, 2014). These will not be reviewed in detail again here, as an excellent review has been provided earlier by Pockett (2013) earlier. Reasons why Libet's Mind Field Theory of consciousness, carved out in his book *Mind Time* (2004), may be discussed outside rather than within the realm of the other field theories are clarified herein. Neural field theories of consciousness, whether they relate to representational fields, where *Gestalten* or *qualia* are seen as reflecting the very nature of consciousness, occupying a presumed spatio-temporal brain field generating electrical brain states (Köhler, 1940), or to the functionally specific spatio-temporal structure of an electromagnetic field in the brain (Lashley et al., 1951, McFadden, 2002; John, 2002; Fingelkurts and Fingelkurts, 2002), all account for specific aspects of brain-behavior function while humans are in a conscious or non-conscious state. Yet, consciousness is a far more complex product of brain evolution, both at the phylogenic (Cabanac, Cabanac, and Parent, 2009) and at the ontogenetic (Jaynes, 1990; Feinberg and Mallatt, 2013) scales, that reaches well beyond brain-behavior function (e.g. Pockett, 2000). Since the theory of evolution, the problem of the origin of mind or, more specifically, the origin of consciousness has been posed. Where and how in evolution has it begun to emerge, and how can it be measured. How can we derive mindfulness out of mere biophysical matter? If we want to define consciousness adequately, it would need to be in terms of the capability of the human Self to know and analyze its own condition and existence in space and time, and to project this knowledge into a future that has not yet happened. Why it may not be possible to render the whole complexity of the phenomenon of consciousness scientifically operational is discussed further in the following chapters. Ontological links between mind, time, and the Self as a window to understanding a specific aspect of human consciousness, the ability to project one's own existence into the future, are brought forward. Finally, the question whether we need neural field theories of consciousness as an explanation of the latter at all is justified under the light of clear argumentation. It is concluded that investigating functional links between eudaimonic well-being and consciousness could give us answers more important to the future of humankind.

1. Limitations to a scientifically operational definition of consciousness

None of the field theories of consciousness has succeeded in providing a definition that would be both scientifically operational and, at the same time, capture the complex nature of this phenomenon. This was already pointed out some time ago by Block (2007) as a clear limitation to not only field theories, but any theory of consciousness. He argued for an "abstract solution" to the "problem of consciousness" given that phenomenal consciousness by far exceeds perception, cognitive accessibility and performance, or any of their directly measurable brain correlates. What others have referred to as the "hard problem of consciousness" (e.g. Chalmers, 1996; Searle, 1998) relates to the difficulty of finding brain measures of a highly complex phenomenon, the conscious *Self*, experienced in terms of *I do*, *I think*, *I feel, I was, am, and will be*, independently of any particular conscious perception, memory, decision, or action (behavior). If a representational or neural field of consciousness occupying a presumed space-time continuum inside the brain, or outside the brain, as suggested by Sheldrake (2013) and others, exists, it would have to be independent of the neural activities underlying any particular perceptual or cognitive process operating at the same time. While a specific conscious perception or conscious memory recall, can be measurably correlated with specific brain activities (e.g. Nani et al., 2019), interpreted adequately as the neural correlates of the particular conscious behavior being measured, it is not a neural correlate of consciousness. The contents of phenomenal experience are associated with neural activities in multiple networks (Rees, Kreiman, and Koch, 2002) of the temporal-parietal-occipital brain areas, but this does not account for how and why, or through which mechanisms, our brain has evolved consciousness, while most other mammal brains have not, or at least not to the same extent. To get around this problem, the concept of a "conscious brain state", with an abstract and scientifically operational definition of consciousness in terms of a "continuous process with limited duration" was introduced (Tononi and Edelman, 1998, and subsequently within a larger theoretical framework of consciousness, Edelman, 2003) based on a definition proposed earlier (von der Malsburg, 1997). Such a parsimonious definition of consciousness may allow looking for invariant properties of a brain state during conscious wakefulness and its equivalents. LaBerge *et al.* (1986; 1990) argued that states of lucid dreaming, for example, are equivalent to wakeful conscious states. The invariant properties of what John called "the conscious ground state", which in his theory corresponds to a specific EEG wave pattern that is observed when patients recover from deep anaesthesia, could then be told apart from the subjective phenomenal contents of any particular conscious experience as reflected in a particular overt behavior. In other words, the conscious brain state would have universal properties that can be consistently identified and measured whether we are consciously daydreaming (Singer, 1975; Carver and Humphries, 1981), engaged in abstract analytic thinking (Gilead et al., 2014), in a state of lucid dreaming (LaBerge, 1990), or whether we consciously perceive and remember objects, as in conscious three-dimensional perception and selective memory recall (Nani et al., 2019). Invariant or universal properties of a generic conscious brain state were, however, never found elsewhere than in brain patterns measured after recovery from deep anaesthesia (John, 2002), which is, yet again, only a particular conscious state to which consciousness is not reducible. Fifteen years ago, the nuclear physicist Jean Durup and myself (Dresp-Langley and Durup, 2006) suggested a biophysical brain code independent of particular cognitive processes operating while we are conscious (perception, spatial awareness, conscious motor planning and execution etc.), but providing a generic neural mechanism that would trigger, maintain, and terminate an individual conscious state in similar ways in which electronic bar codes may activate, maintain, and disable the electronic locks of a safe. The functional assumptions underlying this somewhat wild concept were inspired by earlier work on an arbitrary, and possibly genetically prewired, selection of dedicated neural circuitry for consciousness (Helekar, 1999), temporal coincidence coding in the brain (e.g. Ainsworth et al., 2012), and adaptive resonance theory (Grossberg, 1999; 2000). Our idea then was to carve out a computational hypothesis for conscious in terms of a purely temporal (time bin) resonance of memory signals in long-range neural circuits well beyond functionally identified sensory areas. Such circuitry would first be arbitrarily selected and short-term potentiated by Hebbian synaptic learning, then subsequently consolidated during ontogenesis, long-term potentiated and dedicated to the generation of temporal signatures (firing patterns) of conscious states. Since consciousness is fully operational in the absence of spatially defined stimulus input from the outside world (when we have our eyes closed and daydream, for example, or in lucid dreaming), we daringly proposed that the temporal signatures of consciousness generated in dedicated neural circuits could become progressively de-correlated from spatial signals during ontogenesis. This hypothesis negates the concept of a spatially defined activity field within the brain as a potential correlate of consciousness, and it makes sense under the light of the fact that the phenomenon itself has no measurable spatial dimensionality. Brain activity patterns are representations, as adequately and parsimoniously defined earlier by Churchland (2002), in terms of patterns of activity across groups of neurons which carry information. Interpretations of such patterns in terms of functionally specific correlates (correlate=co-occurring with) of consciousness under specific conditions of testing (for a recent review, see for example, Koch, Massimini, Boly, and Tononi, 2016) all carefully avoid the term causality, as correlation does not necessarily imply causality. Thus, when it comes to an *explanation* of consciousness, we are still found wanting, consistently faced with the same old problem, over and over again. It has up to now not been possible to confirm that specific brain activity patterns, or synchronization thereof, recorded during a specific conscious experience, explains consciousness, or even leads to an understanding of the phenomenon. This is so, because we do not know beyond reasonable doubt whether the brain activity patterns demonstrated in any of the relevant studies in the field are neural signatures of consciousness, or nothing more (or less) than the traces of different levels of integrated brain activity (see also the review by Pockett, 2013), representing ongoing sensations, memories, or mental images during conscious experience.

## 2. Absence of a spatial dimensionality of consciousness

Consciousness has no observable spatial dimensionality (e.g. Libet 2004; Buzsáki, 2007; Marchetti, 2014). It corresponds to a specific state of being in time independent of spatial location. It is a phenomenon that can be observed and accounted for scientifically across changes with time in ontogenetic development. The ability of human beings to both consciously relive past events and conceive future events entails an active process of construction of consciousness in time that underpins many other important aspects of conscious human life. During early childhood, the brain learns about the temporal order of the physical world, well before we become phenomenally conscious of a *Self* and its immediate or distant environments (Piaget, 1967). The temporal rhythms or order of stimuli is the first way through which humans acquire knowledge of physical reality, structure, and continuity, as illustrated by results from experiments where the responses of newborns exposed to speech stream inputs have been systematically quantified (Bulf, Johnson, and Valenza, 2011). Human awareness of temporality and the projection of the *Self* and its most abstract concepts and reasoning towards a future that has not yet happened occurred relatively late in evolution as a mental capacity allowing to push the technological and cultural development of our species further and further. It develops ontogenetically over the first years of an individual lifespan (Piaget, 1967; Jaynes, 1990; Edelman, 1993). The idea of ontological identity link between consciousness and awareness of temporality harks back to Hume (1740), Hegel (1807) and to Heidegger's (1927) concepts of *Sein* (Being) and *Zeit* (Time), where consciousness is hardly more than a succession of psychological moments where we realize that we exist in, and are part of, moments in time. This places all other perceptual or sensorial processes which may characterize any particular phenomenal experience at a different level of analysis. The idea of a fundamental identity link between awareness of Self (*das Ich*) and awareness of what Heidegger termed *Ursprüngliche Zeit* (original time) implies that human consciousness may have progressively evolved from the primitive ability to be aware of, to remember, and to predict temporal order and change in nature found in other species such as rodents (e.g. Fouquet, Tobin, and Rondi-Reig, 2010). The limits of this capacity are directly determined by brain capacity and extent of interaction between the brain and the world. The activity of individual neurons in the brain only has a small quantifiable relationship to sensory representations and motor outputs. The most recent evidence (Ainsworth *et al.*, 2019) confirms that coactivation of a few 10s to 100s of neurons can code sensory inputs and behavioral task performance within clear psychophysical limits. However, in a sea of sensory inputs with memory representations linked to complex motor output, the temporal activity of neurons has to be functionally organized independent of spatial codes relative to sensory data and representations. In the brain, this could be made possible either through spike rates (rate coding), or the selective reinforcement of spike coincidences active neurons and networks (temporal coding). Both have computational advantages and are not mutually exclusive. There is recent evidence (Ainsworth *et al.*, 2019) for a bias in neuronal circuits toward temporal coding. Just as the temporal signal sequence or activity pattern of any single coding cell is determined by its firing activity across a certain length of time, the temporal signature of a conscious state is also linked to duration, with variations in the limited dynamic range of a few hundreds of milliseconds. Most perceptual and cognitive contents are processed implicitly by the brain (i.e. non-consciously) and truly conscious states seem to be reflected by short oscillatory activity periods of not more than a few hundreds of milliseconds each (e.g. Buzsáki, 2007, Del Cul et al., 2009, Nybegrg et al., 2010). The clockworks of consciousness have thus been conceived in terms of rapid temporal successions of microscopic brain states Edelman, 2003). Spike time-based models of processing relating to such have been suggested (e.g. Singer, 2000; Thorpe, Delorme, and van Rullen, 2001). The temporal characteristics of resonant brain networks, under the hypothesis of a functional separation from spatial mechanisms, can explain the temporal stability of representations as defined by Churchland (2002; see here above) despite the highly plastic and diffuse functional anatomy of the brain (Wall, Xu, and Wang, 2002). Such stability is ensured by the dynamics of bottom-up patterns triggering neocortical excitatory activity and matched in time with top-down memory representations or expectation signals (Grossberg, 1999; Dresp-Langley and Durup, 2009; Cacha and Poznanski, 2014). The temporal dynamics of these brain activity patterns are driven by environmental pressure (relevance) towards functional interaction (e.g. Grossberg, 1999; Hari, 2017). Consciousness can also be understood in terms of changes that have occurred in time during phylogenesis. All vertebrates appear to be phenomenally and affectively conscious, function according to circadian cycles, and experience various states of being with dynamic transitions between different states of consciousness (Birch and Schnell, 2020). In humans, however, the variety of higher-order states of consciousness described in the literature is not only larger (Fabbro et al., 2015), but also qualitatively different. So-called primary consciousness related to representations coding for perception, affect, and action (Edelman, 2002) are likely to be shared by all mammals, while higher-order consciousness linked to the interpretation of contents of primary consciousness, self-related awareness of past and future, and symbolic activities relating to language (Kotchoubey, 2018) is what makes humans unique. There is no doubt that conscious ability results from neural network development and processing resources within the physical boundaries of a brain. However, human consciousness transcends these underlying brain mechanisms. By changing and developing with experience, in interaction with other beings, society, and the physical world, it creates potential for mobilizing new resources beyond currently existing or known physical boundaries.

### 3. Consciousness transcends brain, space, and time

The capability of consciously shaping our lives within the world and of projecting them into a distant future, imagined but not yet real, is a critical aspect of fully evolved human consciousness, and drives human creativity and the technological and cultural development of human societies (e.g. Fabbro et al., 2015). As pointed out earlier (Logan, 2007), when humans started living in complex communities where cooperation was a key to survival, through the maintenance of the camp fire among other things, consciousness may have emerged from a new complexity of human interaction. Consciousness may thus be seen as a specific form of energy, with a creative potential that can lead to directly observable changes, in other beings and in physical environments. Consciousness thereby has, or can have, the power to determine and/or change future non-physical (mental) states, in the *Self* and in others, and/or future physical states in the outside world. To clarify how we may link this form of energy to the brain on the one hand, and to human society and the physical world on the other, we may consider the following general definition:

*"Energy is the capacity for doing work. It may exist in potential, kinetic, thermal, electrical, chemical, nuclear, or other forms" (Encyclopedia Britannica, 2021)*

The origin of conscious energy is definitely the brain, its form potential, and the work it does when operational is the work of producing change, in the *Self*, in other people, and in the world. Consciousness is thereby defined as the energy source of all change and creativity. The latter can be as diverse as the things we may observe, in ourselves, in others, and in the world. Consciousness enabled human creativity breaks currently known mental and physical barriers every day in art and science, finding new solutions to problems that previously appeared insurmountable. Phenomena such as consciously guided brain-to-brain communication (Grau et al, 2014) are now being investigated in respectable research laboratories. New particles that do not seem to obey conventional laws of physics have been discovered (Bazavov et al, 2014). Yet, our knowledge about the processes underlying this individual and collective power of consciousness is still very limited. Physical theory entails that the functioning of organisms or environments (living systems) is continuously challenged by the laws of thermodynamics in its attempt to maintain energy. Living organisms interact with each other and their environments in a continuously evolving process aimed at precisely that goal. With the evolution of the species, such interactions become increasingly sophisticated and effective. It is possible that consciousness, well beyond mere biological adaptive function (e.g. Jordan and Ghin, 2006; Kotchoubey, 2018) has evolved for the purpose of extending human capacity to, not only maintain, but create new energy. Interactionism thus becomes a key to understanding consciousness in terms of a product of the "transcendental human brain" (Cacha and Poznanski, 2014) or, in other words, a specific kind of temporally defined energy that results from interaction with space, but is not spatially defined or limited. As pointed out elsewhere (Pepperell, 2018), if we want to explain consciousness as a physical process we must acknowledge that, as in all physical processes, energetic activity is fundamental also to the processes that drive evolution and have produced consciousness. The nature of energy itself can take many different forms in physics. In the case of brain function, energy may be conceived in terms of the forces that produce specific electric activity dynamics (observable as brain waves) while we are conscious. By observing physical systems, we can understand how energy produces temporal change in systemic states on the basis of the observed differences between these systemic states. Consciousness as a transcendental form of energy, produced in the brain but not limited to the latter, produces temporal change in states of the *Self*, as suggested by current insights from studies on deep meditation. Meditative practice detunes the brain processes of self-awareness and blocks the instantiation of self-referential conscious states (Nair et al., 2017; Keppler, 2018). This leads to what has been called the "dissolution of the ego", which is not to be confounded with dissolution of consciousness (Keppler, 2018). Instead, deep meditative states tap into a wider spectrum of functional brain modes, opening the gates to extended phenomenal experience and expanded consciousness. The subject/object relationship, or relationship of the Self to its immediate environment during such a transcendental experience (Nair et al., 2017) is characterized by absence of a conscious perception of time passing, space, or body sense, in other words anything that would give meaning to conscious waking experience. Transcendental consciousness is described as full self-awareness isolated from the processes and objects of experience but characterized by the absence of sense of time, space, the physical body or the contents of perception that define waking experiences (Vieten et al., 2018; Paoletti and Ben-Soussan, 2020). The integration of transcendental experience with waking, dreaming, and sleeping has been labeled by distinct subjective and objective markers. It is subjectively marked by full self-awareness during waking, sleeping, or dreaming, greater emotional stability, decreased anxiety during challenging tasks and, physiologically, by the coexistence of specific brain wave patterns (Travis, 2014). Transcendental experience may be an engine that fuels human development and creates potential for the evolution of higher forms of human "intelligence", to be understood here in terms of mental capability. Insights from the neuroscience of meditation and its effects on human well-being and the development of higher forms of consciousness (Muehsam et al., 2017; Mahone et al., 2018; Vieten et al., 2018; Brandmeyer, Delorme, and Wahbeh, 2019; Vivot et al., 2020) points towards states of enhanced consciousness in deep meditation as a crucial driver of human psychological development. Deep meditation also has proven therapeutic effects, as in chronic pain management (Hilton, Hempel, and Ewing, 2017) in individuals where consciousness is often reduced to little else than overwhelming sensations of pain, limiting the full, health expression of conscious capability in their everyday lives. One of the most controversial issues in vegetative state or a minimally conscious state syndromes concerns the potential capacity of such patients to continue to experience pain in the absence of any measurable self- or environmental awareness. Some of such patients might continue to experience elementary emotions or feelings, as suggested by results from neuroimaging studies showing activation of specific cerebral areas in response to situations which commonly generate empathy (Pistoia et al., 2013). In short, there are still more questions to be answered in consciousness research, well beyond what we call biological adaptation in the Darwinian tradition. For example, the intimate link between transcendental states of mind and a phenomenon called *eudaimonia* needs to be unraveled in well-targeted research across the human lifespan. This novel, largely uncharted terrain of scientific investigation into human consciousness may provide deeper insights into its function far more important to our species than biophysical explanations in terms of neural correlates, or biophysical fields. As pointed out by Churchland (2005), we have to, ultimately, ask ourselves what we want from a science of consciousness.

**4.   Eudaimonic well-being as a window to the purpose of consciousness**

*Eudaimonia* is a central concept in the Aristotelian philosophy of ethics. It is related to other concepts such as virtue, human excellence, and *phronesis*, which is an ethically grounded form of wisdom (cf. Walsh, 2015). The viewpoint on consciousness delivered here suggests potential for generating creative change, in non-physical (mental) states of the *Self* , of others, or in physical states as a neglected possibility for scientific investigation. The newly emerging field of research on eudamonic development (e.g. Rosenfeld, 2019; Alexander et al., 2021) can be linked to the clinical neuroscience of expanded consciousness (Paulson et al., 2017; Vieten et al., 2018; Vivot et al. , 2020) to generate new insights on how we can foster acceptance of what cannot be changed, in ourselves and in others. It could help us understand how expanding our consciousness can help us adjust our individual expectations, and find purpose and fulfillment despite adversity. A deeper understanding of consciousness in terms of practice (mindfulness, deep meditation) could help us understand the intuitive processes that enable us to think "outside the box",  and to transcend the limitations of preconceived knowledge. Unconscious mechanism play an essential role in these intuitive processes (e.g. Grobstein, 2005; Dresp-Langley, 2011; Paulson et al. 2017b). Eudaimonia most certainly fuels on consciousness as energy potential, as a fundamental driving force towards a better life, referring to the subjective experiences associated with living a life of virtue and purpose in the pursuit of human excellence. The clinical neurosciences have only just begun to explore the underlying psychological and physiological mechanisms and processes. The phenomenological experiences derived from eudaimonic living include self-realization, self-actualization, personal expressiveness, optimism, vitality and, as experienced in states of deep meditation, capability of transcending the Self (Di Fabio and Palazzeschi, 2005). While hedonic well-being focuses on happiness defined in terms of pleasure and the avoidance of pain, the eudaimonic approach, on the other hand, relates to meaning and self-realization where well-being is seen as the full functioning of a person at a higher of consciousness than the one that seeks pleasure to attain happiness. A conscious Self in a state of eudaimonic well-being is focused on inner resources, resilience, higher meaning, authenticity, and purposefulness (Cosco, Howse, and Brayne, 2017; Di Fabio and Palazzeschi, 2015). Consciousness changes as our brains age, with changes in challenges to meet, and in perspectives for the future. Although younger individuals may be deemed more resilient than ageing ones, this is a preconception that needs to be reconsidered in the current society context under the light of new challenges and problems related to current society contexts, the internet, and social media (e.g. Garland, 2018; Dresp-Langley, 2020; Liu, Jiang, and Zhang, 2021). In ageing individuals, on the other hand, changes in embodiment occur with ageing neuronal mechanisms and their associated sensorimotor and cognitive deficits (Costello and Bloesch, 2017). Investigating the affective and cognitive mechanisms of deep meditation, mindfulness, and expanded consciousness at the behavioral, metabolic, and neurobiological levels, with research into the mechanisms of action underlying expanded consciousness, will no doubt contribute to the development of treatment options beyond age-related changes in neuronal function. A wider range of comprehensive approaches needs to be developed to address the multiple mechanisms underlying the many different conditions of breakdown of human resilience (e.g. Franklin, 2012). Such may be, but not always, related to brain aging. While embodiment changes as we age (Kuehn et al., 2018;), these changes do not inevitably produce lesser well-being (Ryff et al., 2016). Yet, bodily awareness is a central component of human consciousness (Treves et al., 2019), and clearly an important aspect of well-being while we are young (Ginzburg et al., 2014). Resilience theory (e.g. Maltby et al., 2019) reflects the general idea that managing to navigate adversity and maintaining high levels of functioning across a large number of various domains demonstrates resilience, i.e. the capacity to cope with adversity. Similarly, traditional models of healthy ageing (cf. Wong, 2018, for a review) suggest that having a high level of functioning across a large number of various domains would be a requirement for well-being and resilience. Yet, this may not necessarily be so. For example, health benefits have been identified among older adults who maintain a purposeful life engagement and thereby exhibit a high level of eudaimonic well-being (Cosco, Howse, and Brayne, 2017). These benefits would include extended longevity, reduced risks in various disease outcomes, reduced physiological deregulation, and gene expression linked to better inflammatory profiles (Ryff et al., 2016). Similarly, the brains of mindful or meditating individuals may be less affected by ageing processes under the light of research suggesting that meditation and similar forms of mindfulness practice, or the states of expanded consciousness such practice generates, could contribute to preserving healthy brain tissue, cognitive and emotional resilience, and diminish the risk of dementia and other age-related neurodegenerative diseases (Kurth, Cherbuin, and Luders, 2017; Lutz et al., 2018). However, as already pointed out herein, the capability of projecting its own existence into a future that has not happened yet is the sole unique property of a fully conscious Self. Only the human primate has evolved this capability and, as postulated here, this capability is intimately linked to eudaimonic well-being as defined here above. This working assumption stems from the, previously discussed (Dresp-Langley, 2011, 2018; Dresp-Langley and Durup, 2012), ontological link between consciousness and time itself, which is summarized again in the context of this paper here, for illustration, in Figure 1. In the current societal context, with the many challenges we have to face and to anticipate, expanding our consciousness and exploiting it mindfully could be the key to our future as well as that of the whole planet. One of the most important functions of expanded consciousness may be its power to generate holistic resilience, to increase our potential to cope with new forms of adversity, whether we are young or old. A science of consciousness that leads to novel forms of well-being in the face of increasing adversity could have a significant impact, for individuals and for society as a whole.

5. **Conclusions and perspectives**

Field theories of consciousness where the latter is seen as having duration and extension in space are limited by the fact that, unlike in the field theories in physics, particular regions of the presumed space-time continuum and interactions between elementary objects herein cannot be objectively measured, or accounted for mathematically. Libet (1994) was well aware of this fundamental problem by recognizing that "the mind field of consciousness" does not correspond to any category of known physical fields and cannot be observed directly by known physical means. Pockett (2013) wrote that "a field that is not observable directly by known physical means is in some danger of remaining confined to the realms of philosophy", leaving it to her readers to decide whether this statement is to be considered as outrageous, or as a sign of a healthy sense of humor. Indeed, what strikes in all of the existing neural theories of consciousness including the field theories is the preconception that there has to be a "scientific" (as opposed to "philosophical") account for consciousness. This is even more astonishing than the "astonishing hypothesis" (Crick, 1994) of a neural correlate for consciousness itself, because it means dismissing the fact that all contemporary science, including mathematics and physics, stems from nothing else but philosophy. Thus, quite ironically, the contemporary science of consciousness is based on the preconception that all reality has to be material in the sense of measurable by the known tools of physics, and that consciousness must be a direct product of physical activity in the brain. Yet, the biggest nut to crack for this kind of materialism has remained the existence of consciousness. It is the latter which has allowed to conceive all our methods and tools for scientific measurement, yet, it looks like these methods currently fail to fully explain what has made their conception possible in the first place. The full recognition that our conscious minds are not confined to our brains, or to what is currently measurable inside the brain or beyond, may eventually free contemporary neuroscience. Why look for a biophysical explanation of consciousness? Pressing society needs call for a science of consciousness in terms of its power to generate and foster eudaimonic well-being, of individuals as well as nations. Consciousness is above anything else an internal, dynamic process that governs intentionality for the purpose of creativity and social interaction and their complex relationships with the first-person perspective (e.g. Metzinger and Gallese, 2003). Thus, consciousness may have evolved to form a "mind field" that reaches beyond space and time to enable us, ultimately, to plan for both our survival, and our extinction (two essential and complementary concepts in Darwin's theory of evolution) when survival on the planet is no longer an acceptable option for our species.

LaBerge S. Lucid dreaming: psychophysiological studies of consciousness during REM sleep. In: Bootzen RR, Kihlstrom JF, Schacter DL, Eds. Sleep and Cognition. 1990;Washington, DC, USA: APA press, pp. 109–126.

LaBerge S, Levitan L, Dement WC. Lucid dreaming: physiological correlates of consciousness during REM sleep. Journal of Mind and Behavior. 1986; 7:251–258.

Lashley KS, Chow KL, Semmes J. An examination of the electric field theory of cerebral integration. Psychological Review. 1951; 58, 123-136.

Libet B. *Mind Time*. Cambridge, Mass, USA: Harvard University Press; 2004.

Liu X, Jiang J, Zhang Y. Effects of Logotherapy-Based Mindfulness Intervention on Internet Addiction among Adolescents during the COVID-19 Pandemic. Iran J Public Health. 2021; 50(4):789-797.

Logan RK. The Extended Mind: The Emergence of Language: The Human Mind and Culture. 2007; University of Toronto Press: Toronto, Canada.

Lutz A, Klimecki OM, Collette F, Poisnel G, Arenaza-Urquijo E, Marchant NL, De La Sayette V, Rauchs G, Salmon E, Vuilleumier P, Frison E, Vivien D, Chételat G; Medit-Ageing Research Group. The Age-Well observational study on expert meditators in the Medit-Ageing European project. Alzheimers Dement (N Y). 2018; 4:756-764.

Mahone MC, Travis F, Gevirtz R, Hubbard D. fMRI during Transcendental Meditation practice, Brain and Cognition. 2018; 123: 30-33.

Maltby J, Day L, Hall SS, Chivers S. The Measurement and Role of Ecological Resilience Systems Theory Across Domain-Specific Outcomes: The Domain-Specific Resilient Systems Scales. Assessment. 2019; 26(8):1444-1461.

Marchetti G. Attention and working memory: two basic mechanisms for constructing temporal experiences. Front Psychol, 2014; 5:880.

McFadden J. The conscious electromagnetic information field theory: the hard problem made easy? Journal of Consciousness Studies. 2002 ; 9(8) : 45-60.

Metzinger T, Gallese V. The emergence of a shared action ontology: building blocks for a theory. Conscious Cogn. 2003; 12(4):549-71.

Muehsam D, Lutgendorf S, Mills PJ, Rickhi B, Chevalier G, Bat N, Chopra D, Gurfein B. The embodied mind: A review on functional genomic and neurological correlates of mind-body therapies. Neuroscience & Biobehavioral Reviews. 2017; 73:165-181.

Nair AK, Sasidharan A, John JP, Mehrotra S, Kutty BM, Just a minute meditation: Rapid voluntary conscious state shifts in long term meditators. Consciousness and Cognition. 2017; 53: 176-184.

**Figure and caption**

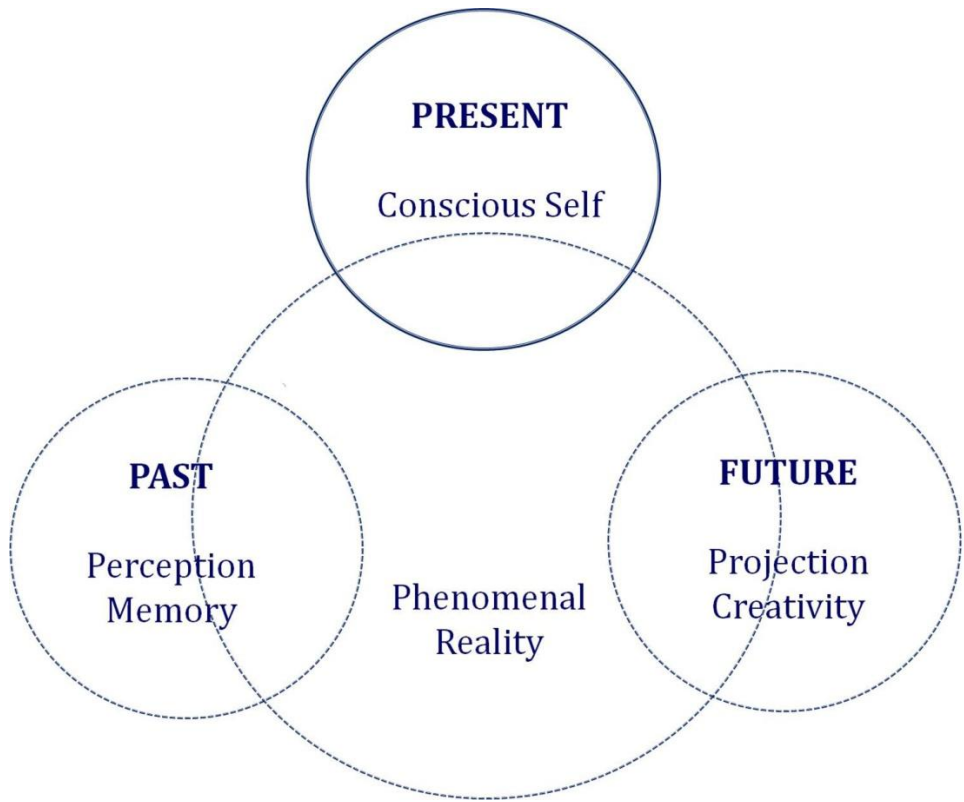

**Figure 1**: All phenomenal reality originates from the ontological link between time and consciousness (after Dresp-Langley and Durup, 2012). This leads to consider consciousness as a form of creative energy beyond space and time, where specific cognitive abilities such as perception, memory or projective thinking and reasoning, although they may exploit conscious energy, need to be placed at a separate ontological level. An optimally expanded level of consciousness (by deep meditation or other mindfulness practice) would be equivalent to an expanded Self in a state of deep of sense of present, past, and future at one and the same moment in time. Such deep states of consciousness may not be attainable by our species.